\documentclass[manuscript]{aastex}

\usepackage{graphicx}
\usepackage{amssymb}
\usepackage{natbib}
\usepackage{amsmath}
\usepackage{url}

\shorttitle{CME acceleration and non-thermal flare characteristics
}
\shortauthors{Berkebile-Stoiser et al.}

\begin{document}

\title{Relation between the CME acceleration and the non-thermal flare characteristics}


\author{S. Berkebile-Stoiser\altaffilmark{1}}
\author{A.M. Veronig \altaffilmark{1}}
\author{B.M. Bein \altaffilmark{1}}
\author{M. Temmer \altaffilmark{1}}

\affil{\altaffilmark{1} Institute of Physics, University of Graz, A-8010 Graz, Austria; asv@igam.uni-graz.at}

\begin{abstract}

We investigate the relationship between the main acceleration phase of coronal mass ejctions (CMEs) and the particle acceleration in the associated flares as evidenced in RHESSI non-thermal X-rays for a set of 37 impulsive flare-CME events. 
Both CME peak velocity and peak acceleration yield distinct correlations with various parameters characterizing the flare-accelerated electron spectra. 
The highest correlation coefficient is obtained for the relation of the CME peak velocity and the total energy in accelerated electrons ($c = 0.85$), supporting the idea that the acceleration of the CME and the particle acceleration in the associated flare draw their energy from a common source, probably magnetic reconnection in the current sheet behind the erupting structure. 
In general, the CME peak velocity shows somewhat higher correlations with the non-thermal flare parameters than the CME peak acceleration, except for the spectral index of the accelerated electron spectrum which yields a higher correlation with the CME peak acceleration ($c \approx -0.6$), 
indicating that the hardness of the flare-accelerated electron spectrum is tightly coupled to the impulsive acceleration process of the rising CME structure.  
We also obtained high correlations between the CME initiation height $h_0$ and the non-thermal flare parameters, with the highest correlation of $h_0$ to the spectral index $\delta$ of flare-accelerated electrons ($c\approx 0.8$). This means that CMEs erupting at low coronal heights, i.e.\ in regions of stronger magnetic fields, are accompanied with flares which are more efficient to accelerate electrons to high energies. 
In the majority of events ($\sim 80\%$), the non-thermal flare emission starts {\it after} the CME acceleration, on average delayed by $\approx 6$~min, in line with the standard flare model, where the rising flux rope stretches the field lines underneath until magnetic reconnection sets in. We find that the current sheet length at the onset of magnetic reconnection is $21 \pm 7$~Mm.
The flare HXR peaks are well synchronized with the peak of the CME acceleration profile, in 75\% of the cases they occur within $\pm5$~min. Our findings provide strong evidence for the tight coupling between the CME dynamics and the particle acceleration in the associated flare in impulsive events, with the total energy in accelerated electrons being closely correlated to the peak velocity (and thus the kinetic energy) of the CME, whereas the number of electrons acclerated to high energies is decisively related to the CME peak acceleration and the height of the pre-eruptive structure.

\end{abstract}

\keywords{Sun: coronal mass ejections (CMEs) --- Sun: flares --- Sun: X-rays, gamma rays}

\section{Introduction}
Coronal mass ejections (CMEs) and solar flares reside among the most powerful and impressing manifestations of solar activity \citep[for overviews see, e.g.,][]{kahler1992,schwenn2006}. CMEs and flares may or may not occur together, 
with the association rate strongly increasing for more energetic events \citep[][]{sheeley1983,yashiro2009}. The question, if and how the two phenomena are physically linked has been widely debated. 
The commonly accepted model of a combined CME-flare event is the eruptive flare scenario \citep[e.g., review by][]{priest2002}. A flux rope embedded into a magnetic arcade starts to rise, causing the magnetic field lines that tie the coronal structure to the solar surface to become more and more stretched to finally form a vertical current sheet beneath the eruption. If the field lines in the current sheet start to reconnect, the sudden release of magnetic energy powers a solar flare. In addition, the newly reconnected field lines 
add poloidal magnetic flux to the rising flux rope, and thus sustain the upward propelling force \citep[e.g.][]{vrsnak2008}. In this way, the energy released in magnetic reconnection is supposed to be distributed both to enhance the kinetic energy of the CME flux rope and to drive dynamic processes in the associated flare, such as the generation of shocks, outflow jets, plasma heating, and the acceleration of high-energetic particles. 

Observational evidence for the coupling and correlation between the flare and the CME characteristics has been presented in several studies dealing with large event samples. Most commonly, such studies use proxies for the energetics of flares and CMEs that can be easily derived from observations, such as the GOES soft X-ray peak flux of the flare and the mean plane-of-sky speed of the CME \citep[e.g.][]{moon2002,burkepile2004,vrsnak2005,mahrous2009}. Recently, also studies of the full CME acceleration profile have been performed, reporting a close synchronization of the impulsive CME acceleration phase and the rise phase of the soft X-ray flux of the associated flare in at least 50\% of the events under study \citep{zhang2001,maricic2007,bein2012}.

Measurements of the CME acceleration are difficult, since the impulsive acceleration of the eruption often lasts only some tens of minutes \citep[e.g.][]{zhang2004,zhang2006} and takes place close to the Sun at distances $\lesssim 3$~$R_\odot$ \citep[e.g.][]{macqueen1983,stcyr1999,vrsnak2001}. This means that imaging of the low corona at high cadence is required. Recent studies have shown that high cadence EUV imagery in combination with white-light coronagraphs provide a good means in order to trace the onset and early stages of CME eruptions \citep[e.g.][]{gallagher2003,vrsnak2007,temmer2008}. 
\cite{bein2011} report that in about 70\% of 96 impulsive CMEs that were studied in such combined EUV and white-light imagery, the CME peak acceleration occured at heights as low as $\lesssim$~0.5~$R_\odot$. 

Important information on the primary energy release in solar flares can be obtained from hard X-ray (HXR) spectra.  Supra-thermal electrons accelerated during the impulsive energy release process precipitate toward the solar surface where they lose their energy in Coulomb collisions with the ambient plasma, heating it to several million degrees. The heated chromospheric plasma expands into the coronal part of the flare loop, where it causes enhanced soft X-ray emission. A tiny part $(\sim 10^{-5})$ of the kinetic energy in non-thermal electrons impinging on the chromosphere is radiated away as non-thermal bremsstrahlung in the HXR domain. This HXR radiation by itself is energetically not important but the spectral characteristics of the radiated bremsstrahlung provide important diagnostics on the energy distribution and the total energy in the flare-accelerated electrons, which contain a large fraction of the total energy released during a flare \cite[e.g.][]{hudson1991,dennis2003}. 

Whereas many studies use observations of the thermal flare plasma, as observed in the soft X-ray domain (primarily by the GOES satellites), to characterize the flare evolution, there are only a few studies which consider the information on flare-accelerated electrons contained in HXR data in comparison with the associated CME dynamics. \citet{qiu2004} and \citet{jing2005} inferred magnetic reconnection rates from the apparent motion of chromospheric flare ribbons and found that the reconnection rate was temporally correlated with the CME/filament acceleration as well as with the flare HXR emission. \citet{temmer2008,temmer2010} presented detailed case studies of the impulsive acceleration in fast CMEs and the evolution of the HXR flux and spectral characteristics of the associated flare, finding a tight synchronization between the flare HXR peak and the CME acceleration peak. However, a study on the relation between the CME acceleration and the evolution of the associated flare energy release and particle acceleration for a larger event sample is still missing. Such study can provide insight not only into the temporal correlation but also into the scaling between characteristic parameters of the flare energy release and the CME acceleration.

In the present paper, we study a sample of 37 impulsive CME-flare pairs for which the CME acceleration phase could be measured and where 
hard X-ray observations of the flare peak were available. The CMEs are observed at high spatial and temporal resolution 
by the EUV imagers and white-light coronagraphs onboard the Solar Terrestrial Relations Observatory \citep[STEREO;][]{kaiser2008}. Using high resolution X-ray spectra provided by the Reuven Ramaty High Energy Solar Spectroscopic Imager \citep[RHESSI;][]{lin2002}, we study the characteristics of the accelerated electron spectra as well as the hot flaring plasma.

\section{Observations}\label{sec:observations}

For the study of the CME kinematics, acceleration and source region characteristics, we used coronal EUV and white light images provided by STEREO's Sun Earth Connection Coronal and Heliospheric Investigation suite \citep[SECCHI;][]{howard2008}. The SECCHI Extreme Ultraviolet Imager \citep[EUVI;][]{wuelser2004} observes the solar disk and off-limb corona up to a distance of 1.7~$R_\odot$ from sun center. EUVI delivers filtergrams in four passbands observing plasma at chromospheric and coronal temperatures. 
We mainly used the 171~{\AA} passband (dominated by emission of Fe~\textsc{ix/x} ions, $T\sim 10^6$~K), and in some cases the 195~{\AA} passband (Fe~\textsc{xii} and Fe~\textsc{xiv} ions; $T\sim 1.5\times 10^6$~K). The nominal time cadence of the 171~{\AA} images is $2.5$~min but can be as high as $\sim75$~s for campaign data. Images taken in the 195~\AA\, passband have a nominal cadence of $10$~min, which has increased to 5~min in 2009. The evolution of the CME further away from the Sun was followed in data from the STEREO COR1 and COR2 coronagraphs \citep{thompson2003}. COR1 has a field-of-view (FOV) from 1.4 to 4 $R_{\odot}$ from Sun center, COR2 from 2.5 to 15~$R_{\odot}$. The observing cadence of the COR1 observations is mainly 5 minutes (but can be up to 20 minutes), the cadence of COR2 
total brightness images is 30 minutes. The overlapping FOVs of the EUVI, COR1 and COR2 instruments enabled us to identify and connect the same CME structure in the observations by the different instruments with high cadence. 

Flare observations were provided by the Reuven Ramaty High Energy Solar Spectroscopic Imager \citep[RHESSI;][]{lin2002} detecting X-ray and $\gamma$-ray emission from the Sun in the energy range 3~keV to 17~MeV. RHESSI is an indirect Fourier imager providing X-ray images at high angular resolution (as good as $\sim2.3''$) and spectroscopy at unsurpassed spectral resolution ($\sim$1~keV below 100~keV).

Since our aim is to compare the evolution of the energy release in solar flares to the characteristics of the associated CME dynamics, we searched for events for which the CME acceleration phase was well observed and the flare impulsive phase was covered by RHESSI observations. We took care not to include flares which were partially occulted by the solar limb. 
In the time period January 2007 to May 2010 (i.e.\ covering the first 3.5 years of STEREO observations), we identified a sample of 37 CME-flare events, which fulfilled these requirements. The GOES flare class distribution of the events selected is GOES class M: 3, C: 16, B: 11, and $\leq$GOES A: 7 events.\footnote{We note that the STEREO mission was launched in solar minimum conditions, and thus the strongest events are missing in our sample.} Out of these, 14 events showed appreciable non-thermal hard X-ray emission. The remaining 23 events showed either weak non-thermal X-ray emission or solely thermally produced X-ray emission.

We note that in our sample we have a selection bias towards impulsive CMEs, i.e. CMEs that have a short main acceleration phase. Due to the distinct anticorrelation of the CME acceleration duration and the CME peak acceleration \citep{zhang2006,bein2011}, this implies also that the involved acceleration values are high. 
The reason for this selection bias is twofold. On the one hand, we aimed to select CMEs where the main acceleration profile could be reconstructed. This tends to exclude events with gradual (i.e. long-duration, almost constant) acceleration. On the other hand, we aimed at comparing the CME acceleration curves with RHESSI observations of the main flare phase. Since RHESSI has a low-Earth orbit (with an orbital period of 96 min), the solar observations are regularly interrupted by eclipses of the satellite. This again tends to exclude gradual, long-duration events.

\section{Methods}\label{sec:methods}

\subsection{CME kinematics and acceleration}

The height-time curves of the selected CMEs were determined by obtaining the position of the leading edge in STEREO EUVI, COR1 and COR2 running difference image sequences. The raw image data were calibrated and processed to improve the visibility of the CME leading edge. First, the images were reduced with the \verb"secchi_prep.pro" routine available in the SSW (SolarSoftWare) tree, which provides for the subtraction of the CCD (Charge Coupled Device) bias, correction for variable exposure time and conversion to physical units. EUVI images were differentially rotated to a common reference time before running difference images were generated. In case of faint CMEs, a normalizing-radial-graded filter \citep[][]{morgan2006} was applied. For COR1 and COR2 observations, a pre-event image was subtracted and a sigma filter was applied to obtain higher contrasts of the transient faint CME structures. For the measurements of the CME evolution, running difference images were constructed by subtracting from each image the image recorded immediately before. If the time cadence of the data was very high or a CME moved very slowly, we rather created difference images out of frames taken further apart in time ($\sim$5--10~min for EUVI data, $\sim$10--20~min for COR1 data). 


The CME kinematics were then derived by following the evolution of the detected CME leading edge along the main propagation direction, starting from the determined CME-flare source region. We developed an algorithm to automatically identify the CME leading edge based on the information that it appears as a bright front with a sharp intensity drop to regions outside the CME
\cite[for details see][]{bein2011}. This algorithm works fine for clear CME fronts but fails for faint ones, in which case we identified the leading edge by visual inspection. 
We note that our CME height measurements are not corrected for projection effects. However, we predominantly selected events where the source region is located close to the solar limb, in order to minimize the influence of projection effects.  For 60\% of our events, the projected radial distance $r$ from Sun center is $\gtrsim 0.8\,R_\odot$, for 85\% of events  $r\gtrsim 0.6\,R_\odot$. 

\begin{figure}[tbp]
\centering
\includegraphics[scale=1.3]{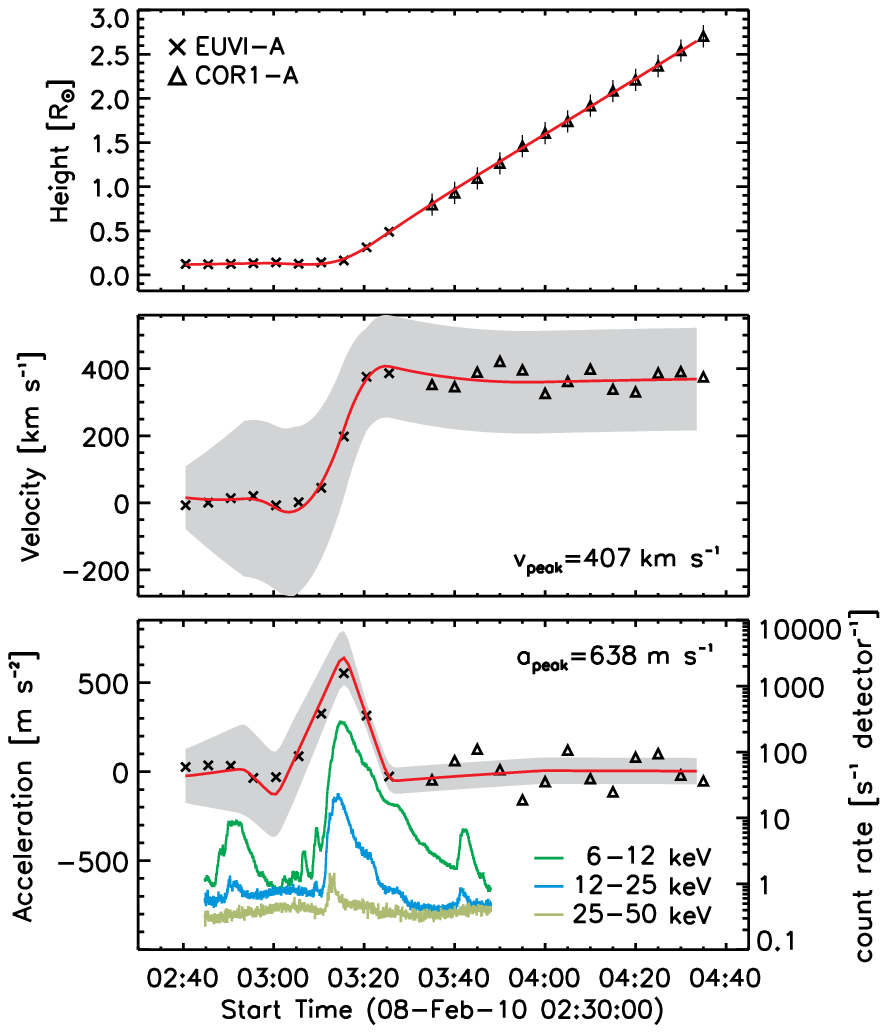}
\caption{Evolution of the C6.2 flare-CME event of 2010 February 8. Top: CME height-time curve. Crosses denote STEREO EUVI, triangles COR1 measurements. 
The line gives the spline fit to the data points. Middle: CME velocity profile derived from numerical differentiation of the height-time curve. 
The grey shaded area marks the estimated errors on the velocity gained from the spline fit (the same applies to the acceleration curve). Bottom: CME acceleration profile and flare hard X-ray emission observed by RHESSI in three energy bands from 6 to 50 keV. (For better visibility, the 6--12~keV and 12--25~keV light curves are multiplied by factors of $0.3$ and $0.1$.)} 
\label{fig1}
\end{figure}
\begin{figure}[tbp]
\centering
\includegraphics[scale=1.]{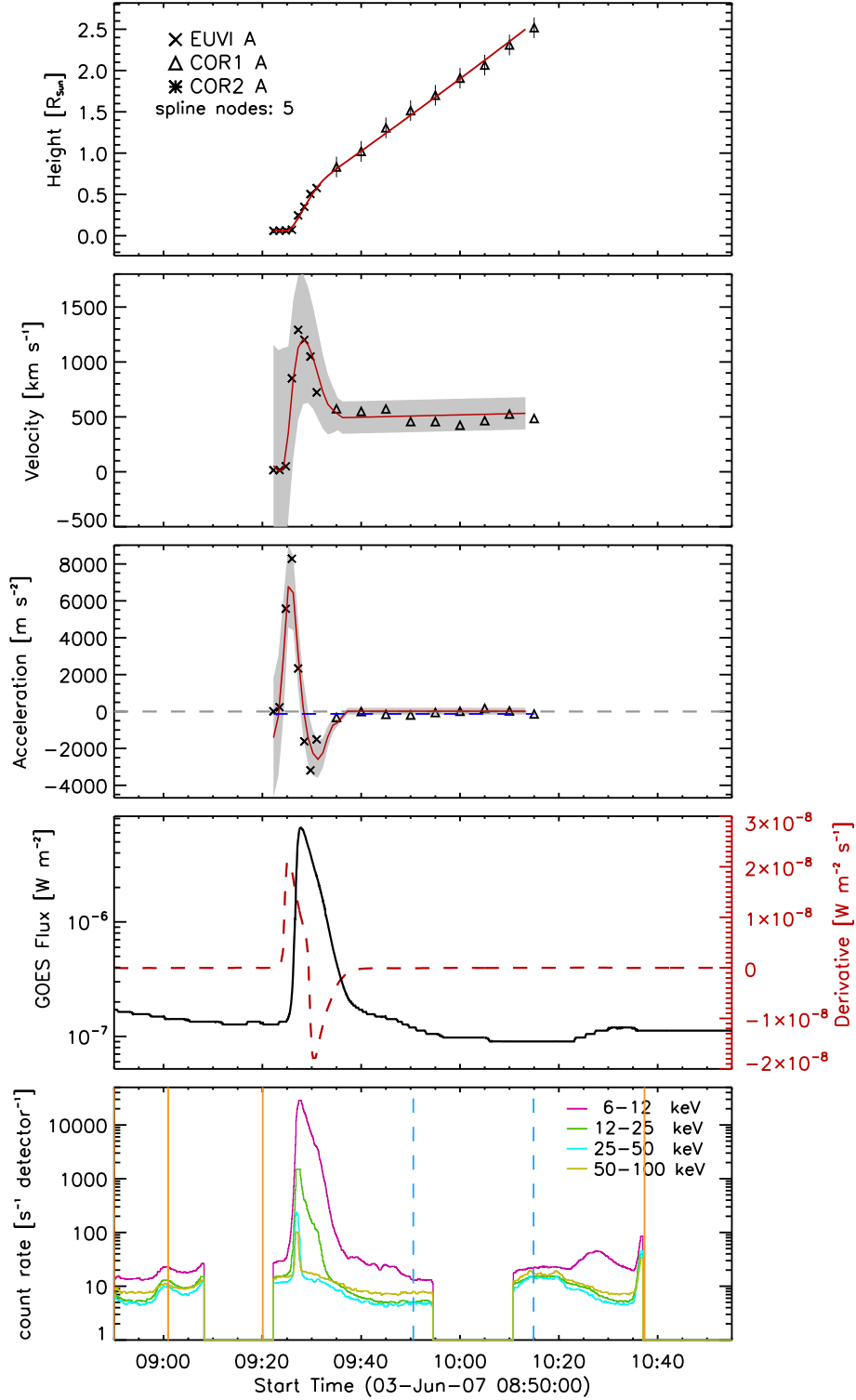}
\caption{CME kinematics and flare X-ray evolution of the C5.3 event of 2007 June 3. From top to bottom:
CME height-time, velocity-time, and acceleration-time curve; flare SXR flux measured by GOES in the 1--8~{\AA} 
band together with its derivative; flare HXR flux observed by RHESSI in four energy bands from 6 to 100 keV.} 
\label{fig2}
\end{figure}
\begin{figure}[tbp]
\centering
\includegraphics[scale=1.]{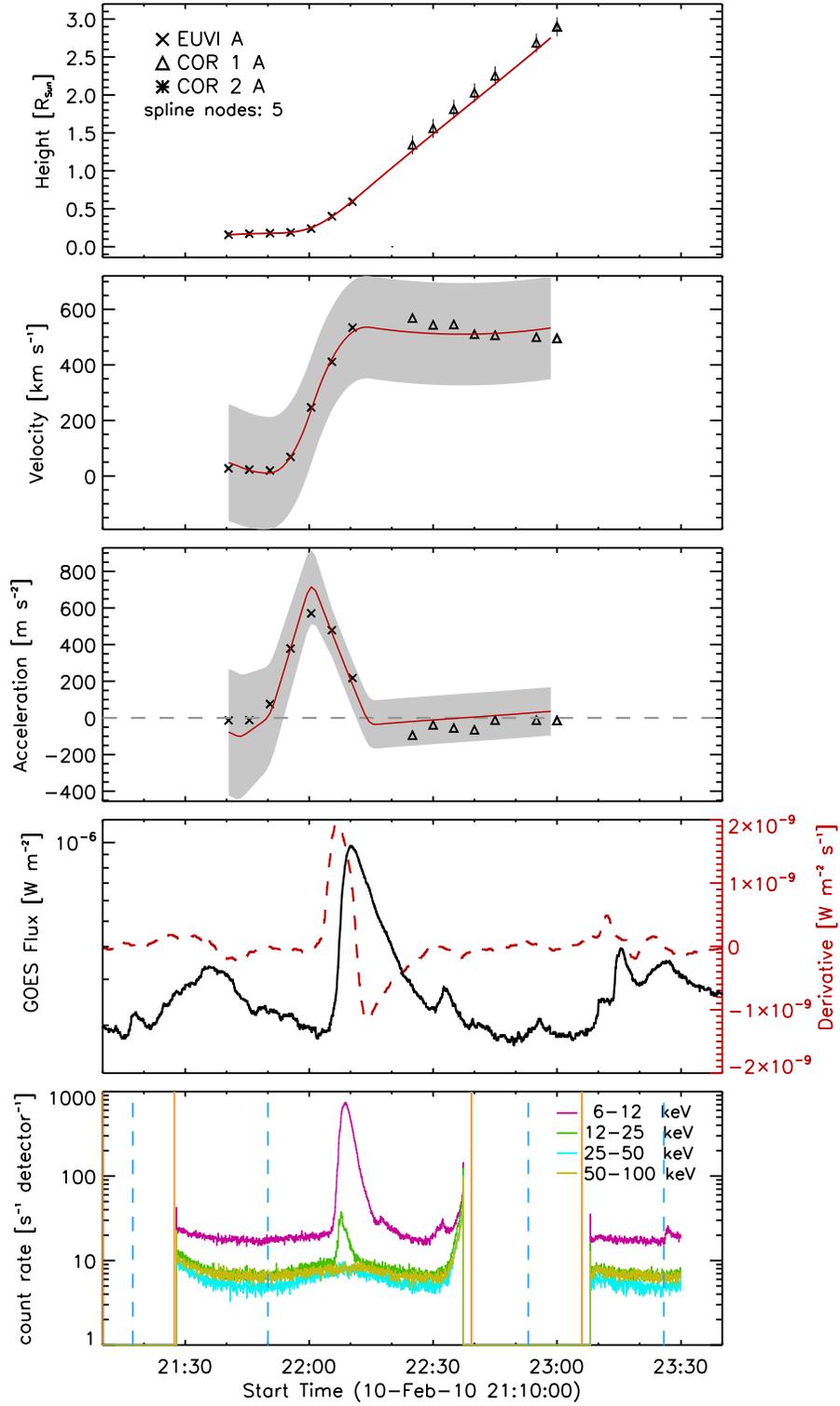}
\caption{Same as in Fig.~\ref{fig2} but for the B9.7 flare-CME event of 2010 February 10.} 
\label{fig3-1}
\end{figure}

Based on the derived CME height-time curves, the velocity and acceleration profiles can be determined by the application of numerical differentiation to the height-time data. Since errors in the height-time curve are enhanced when taking the derivative, a smoothing and fitting method is used based on free-knot cubic splines. This fitting technique also allowed us to estimate errors in CME velocity and acceleration by propagating the uncertainties of the fitted spline coefficients to the first and second derivative. For details on the data processing, automated CME tracking, spline fitting and error analysis we refer to \citet{bein2011}. We note that the 37 CME-flare pairs under study are a subsample of the 95 CMEs that were studied in \citet{bein2011}. In Figure~\ref{fig1}, we show 
the CME height-, velocity- and acceleration-time curves for a sample CME-flare event (2010 February 8) together with the GOES and RHESSI X-ray flux evolution of the associated flare. 
Further examples are shown in Figs.~\ref{fig2} and \ref{fig3-1}.


The parameters obtained from the fitted height-time curves and their first and second derivatives are the CME peak velocity $v_\mathrm{max}$, the CME peak acceleration $a_\mathrm{max}$ and the times at which velocity and acceleration reached their maximum ($t_\mathrm{vmax}$ and $t_\mathrm{amax}$). We also determined the acceleration duration $t_\mathrm{acc}$ defined as the time interval $t_\mathrm{start} < t_\mathrm{amax} < t_\mathrm{end}$, where $t_\mathrm{start}$ and $t_\mathrm{end}$ are the times at which the CME accelerated/decelerated to $\sim10\%$ of its peak value. In addition, we derived characteristic height parameters, namely the height $h_0$ where the CME was first identified, the height $h_\mathrm{vmax}$ at which the CME velocity reached its maximum, and the height $h_\mathrm{amax}$ at which the CME acceleration reached its maximum.
The height $h_0$ of the first CME observation provides us with a rough measure of the CME initiation height and thus also with an upper limit for the size of the pre-eruptive structure \cite[cf.][]{bein2011}.

\subsection{Flare X-ray spectroscopy}

RHESSI X-ray spectra yield information on fast electrons accelerated during the flare process as well as on the thermal flare plasma. For each event, we derived a background-subtracted photon spectrum integrated over 20~s during the hard X-ray peak, i.e.\ the peak of the non-thermal emission, using all RHESSI front detectors except 2 and 7 \citep{smith2002}. In addition, we also derived spectra during the peak of the soft X-ray emission ($\sim$3--12 keV), to better characterize the thermal flare plasma. 

\begin{figure}[tbp]
\centering
\includegraphics[scale=0.8]{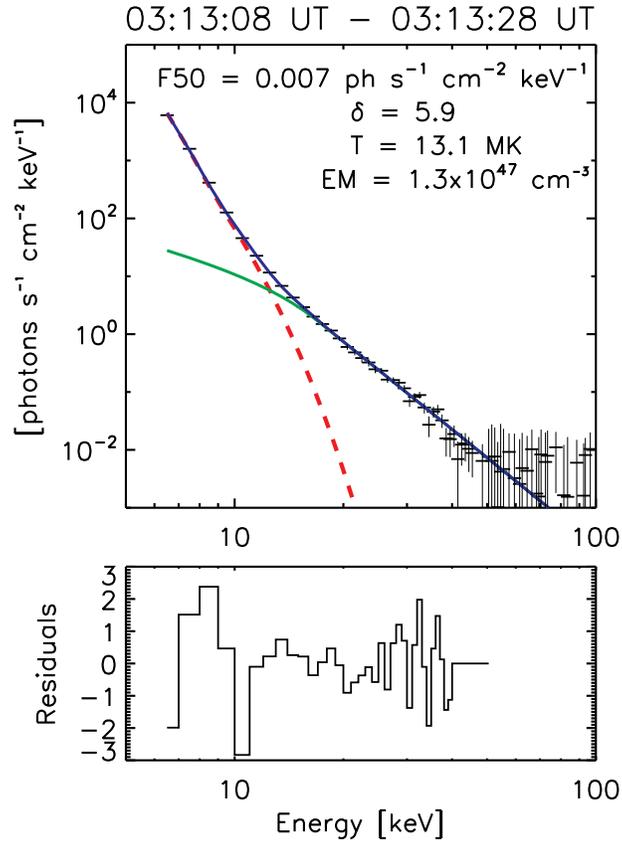}
\caption{Top: RHESSI X-ray spectrum observed at the peak of the GOES C6.2 flare of 2010 February 8 (cf. Fig.~\ref{fig1}). 
The fit to the data points is composed of a thermal (red dashed line) and a non-thermal thick-target bremsstrahlung model (green solid line). The blue line gives the sum of both fit components. Bottom: Normalized residuals to the fit.}\label{fig1a}
\end{figure}

In Figs.~\ref{fig1}--\ref{fig3-1}, we show the CME kinematics together with the X-ray flux of the associated flare for three CME-flare pairs. 
A sample RHESSI spectrum observed at the HXR peak of the C6.2 class flare-CME event on 2010 February 8 is shown in Fig.~\ref{fig1a}. Using the OSPEX software \citep[Object Spectral Executive;][]{schwartz2002}, we applied a forward fit to the spectra using either a combined non-thermal thick-target bremsstrahlung model (at higher energies) and an isothermal model (at the low energy end), or solely an isothermal model. The resulting spectral parameters for the thermal fit comprise the emission measure EM and temperature $T$ of the hot flaring plasma, and for the thick-target model the number of accelerated electrons e$^-$, the electron power-law index $\delta$, and the low-energy cut-off $E_c$ of the accelerated electron spectrum. However, the number of electrons and the low-energy cutoff are intrinsically linked, 
and $E_c$ cannot be determined with accuracy. Thus, as an additional parameter characterizing the strength of the non-thermal emission, we also determined for each power-law spectrum the (fitted) photon flux at 50~keV, $F_{50}$, and the power $P_{20}$ contained in 	electrons accelerated to kinetic energies $>$20~keV.
The obtained flare parameters were then correlated with the parameters characterizing the CME kinematics and dynamics (see Sect.~\ref{sec:vel_acc_profiles}).

In order to study the relative timing of the CME acceleration and the flare energy release as evidenced in the evolution of non-thermal HXR emission, we reconstructed RHESSI light curves at energies above the low-energy cut-off $E_c$ derived from the spectral fits. Based on these light curves we determined start, peak and end time as well as the duration of the non-thermal flare emission, which were then compared with the acceleration profile of the associated CME (see Sect.~\ref{sec:timing}).


\section{Results}\label{sec:results}

\subsection{Correlations of characteristic CME and flare parameters}\label{sec:vel_acc_profiles}

\begin{figure*}[htbp]
\centering
\includegraphics{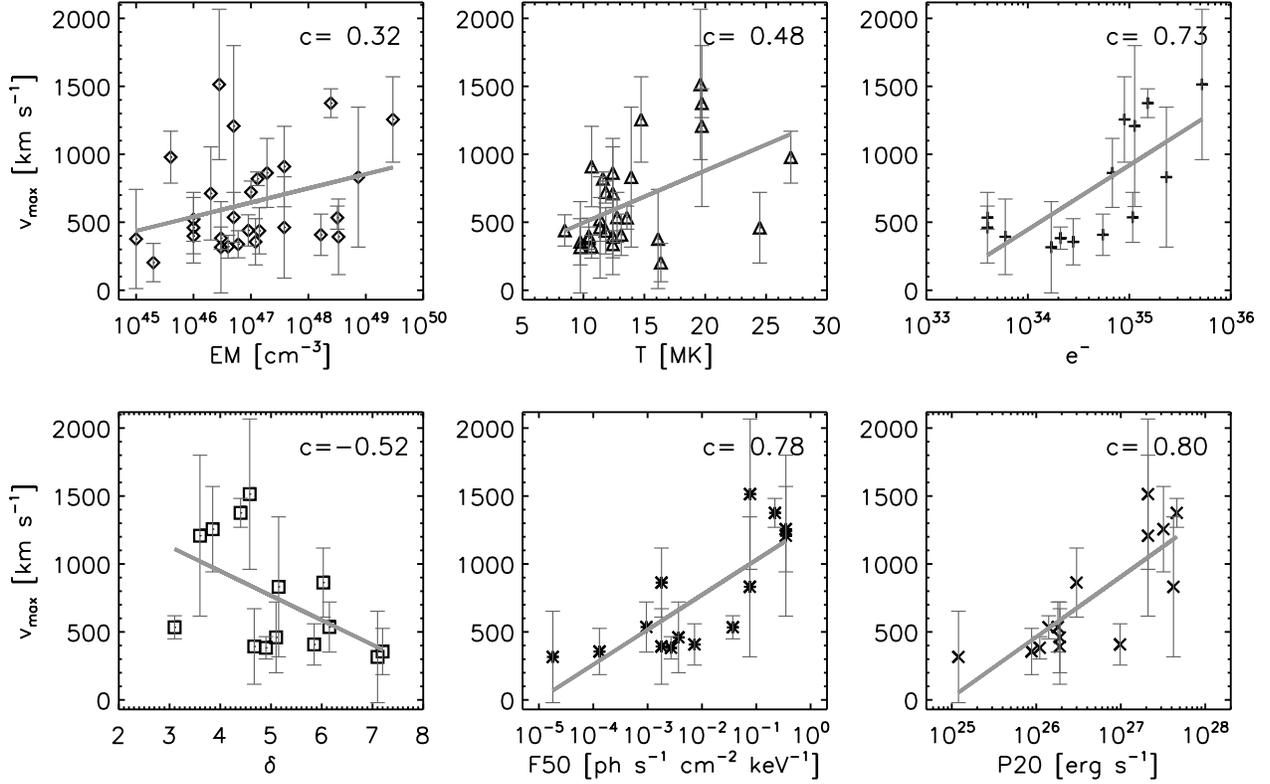}
\caption{Scatter plots of the CME peak velocities against the X-ray spectral parameters of their associated flares, i.e. the emission measure EM, temperature~$T$, number of accelerated electrons~e$^{-}$, hardness of the electron spectrum~$\delta$, 
photon flux $F_{50}$ at 50~keV, and kinetic energy in accelerated electrons with energies above 20~keV. In each panel, we annotate the linear correlation coefficient $c$ and the regression line to the data (solid gray line). 
All flare spectral parameters except $T$ and $\delta$ are plotted on a logarithmic scale.}\label{fig2a}
\end{figure*}

\begin{figure*}[tbp]
\centering
\includegraphics{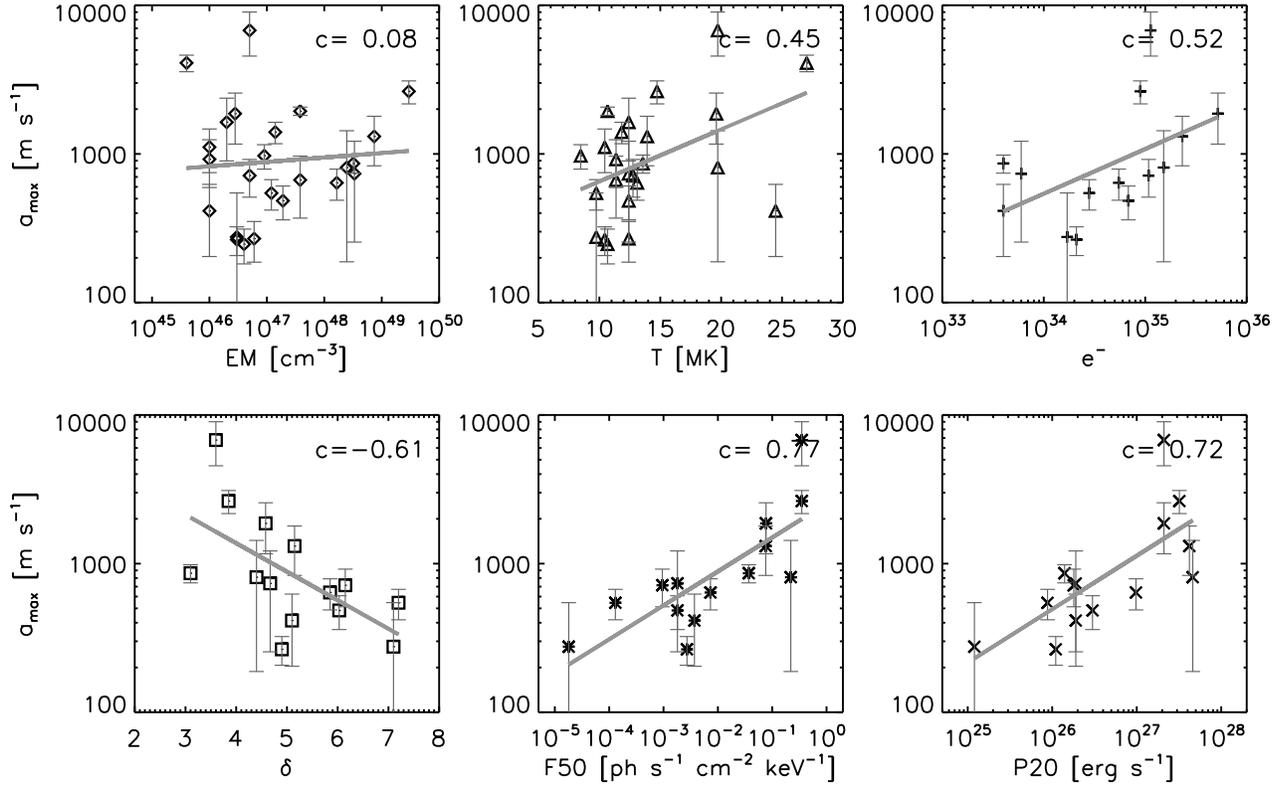}
\caption{Scatter plots of the CME peak accelerations against the X-ray spectral parameters of their associated flares. For further details see the caption of Fig.~\ref{fig2a}. }\label{fig2b}
\end{figure*}

Figs.~\ref{fig2a} and~\ref{fig2b} show scatter plots of the CME peak velocity and CME peak acceleration, respectively, against the characteristic flare spectral parameters, namely the emission measure EM, temperature~$T$, number of accelerated electrons~e$^{-}$, power-law index~$\delta$ of the accelerated electron spectrum, 
photon flux $F_{50}$ at 50~keV, and power $P_{20}$ in accelerated electrons with energies $>$20~keV. Note that the number of data points in the various scatter plots may differ from each other due to the different number of observables available for each CME-flare pair. Our sample covers many weak flares, and not all events show significant non-thermal emission. Thus, we only consider non-thermal fitting parameters for reasonably well observed power-law spectra which have an electron spectral index $\delta\lesssim 8$ (which applies to 14 events out of a total of 37 under study). In each scatter plot, we also show the regression line and the linear correlation coefficient $c$ for the respective quantities. Except one ($EM$ vs. $a_\mathrm{max}$), all the correlations in Figs.~\ref{fig2a} and~\ref{fig2b} are significant at a level of $95$\% or higher.

\begin{figure}[tbp]
\centering
\resizebox{7cm}{!}{\includegraphics{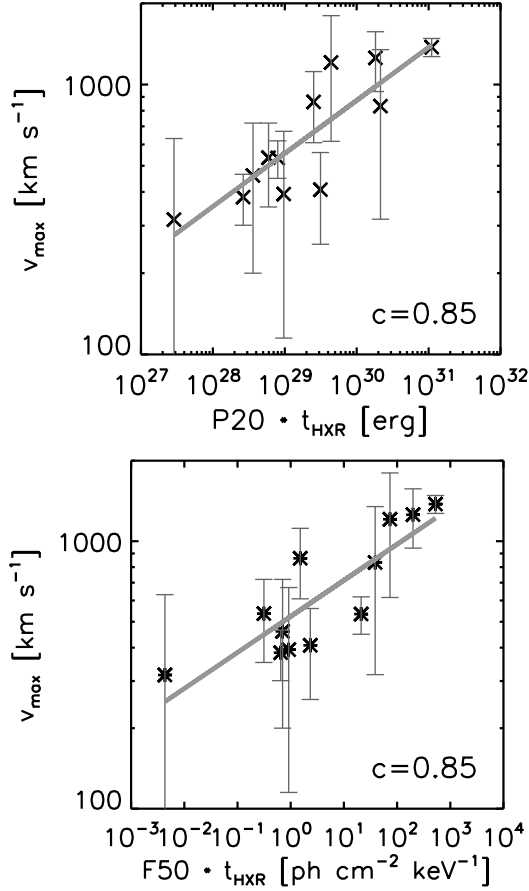}}
\caption{Top: Scaling of the CME peak velocities with an approximation of the energy contained in non-thermal electrons (derived as the product of the power in non-thermal electrons $\gtrsim20$~keV
at the flare peak and the duration of non-thermal HXR emission). Bottom: Same as above but $P20$ is replaced by the non-thermal X-ray flux at 50 keV, $F50$, observed at the flare peak.}\label{fig3}
\end{figure}

Both the CME velocity and the CME acceleration show distinct scalings with the X-ray spectral parameters characterizing the non-thermal energy release in the associated flare. 
The correlation between the CME peak velocity $v_\mathrm{max}$ and the number of flare-accelerated electrons $e^-$ gives a linear correlation coefficient of $c = 0.73$, the relation $v_\mathrm{max}$ vs. $F50$ gives $c = 0.78$, and correlating $v_\mathrm{max}$ and $P20$ we obtain $c = 0.80$ (see Fig.~\ref{fig2b}). In Fig.~\ref{fig3}, we plot the CME peak velocity against the product of the non-thermal power in electrons, $P20$, and the duration of the non-thermal HXR emissions. The correlation coefficient of this product, which is a measure for the total kinetic energy contained in electrons accelerated during the flare impulsive phase, and the CME peak velocity is very high with $c= 0.85$. We obtain the same result when comparing $v_\mathrm{max}$ with the product of the non-thermal photon flux $F50$ and the HXR duration ($c = 0.85$). Thus, the best observed scaling is the flare non-thermal energy and the peak velocity attained in the CME, which is directly linked to its kinetic energy $E = mv^2/2$, 
where $m$ is the total CME mass. The slope of the non-thermal power-law index $\delta$ is found to be inversely correlated with the CME peak velocity, $c = -0.52$, i.e.\ fast CMEs are preferentially associated with flares with harder power-law spectra. We also found a positive scaling of $v_\mathrm{max}$ with the observed duration of HXR emission, $c = 0.58$, i.e.\ CMEs which reach higher velocities tend to be associated with flares of prolonged electron acceleration.

The CME peak acceleration $a_\mathrm{max}$ (Fig.~\ref{fig2b}) reveals correlations with the non-thermal flare parameters comparable to that obtained for $v_\mathrm{max}$ (Fig.~\ref{fig2a}). However, in general the obtained correlation coefficients for $a_\mathrm{max}$ are slightly lower than that obtained for $v_\mathrm{max}$, except for the relation $a_\mathrm{max}$ vs. $\delta$ which is higher, $c = -0.61$. The correlation coefficient of CME peak acceleration $a_\mathrm{max}$ and the number of flare-accelerated electrons $e^-$ is $c= 0.52$. The non-thermal photon flux $F50$ and the electron power $P20$ show a distinct positive scaling with the CME peak acceleration $a_\mathrm{max}$ with correlation coefficients of $c = 0.77$ and $c = 0.72$, respectively. These results are indicative of a tight coupling between particle acceleration in flares and the associated CME dynamics. We can speculate that the fact that $a_\mathrm{max}$ scales somewhat better with the electron spectral index $\delta$ than $v_\mathrm{max}$ implies that the CME peak acceleration is stronger linked to the hardness of the flare-accelerated electron spectrum, and thus with the number of electrons accelerated to high energies, whereas the CME peak velocity is better related to the total number and energy in flare-accelerated electrons, which are dominated by the low-energy end of the particle distribution.

The CME peak velocity and peak acceleration also show a positive scaling with the thermal flare parameters, i.e.\ the emission measure EM and temperature $T$. However, the correlations are significantly smaller than that obtained for the non-thermal flare parameters. The CME peak velocity $v_\mathrm{max}$ and the flare emission measure EM derived at the HXR peak time are weakly correlated with $c = 0.32$, whereas EM and $a_\mathrm{max}$ are basically uncorrelated ($c = 0.08$). A better scaling is observed for the flare temperature $T$ and CME velocity and acceleration with a correlation coefficient of $c = 0.48$ for $v_\mathrm{max}$ vs.\ $T$ and $c= 0.45$ for $a_\mathrm{max}$ vs. $T$. However, the emission measure and temperature derived at the peak of the flare SXR emission are most probably a better indicator of the maximum thermal energy reached in the flare. Indeed, RHESSI EM and $T$ derived at the flare SXR peak show a somewhat better correlation with the CME $v_\mathrm{max}$ and $a_\mathrm{max}$, with the highest correlation coefficient of $c\approx 0.5$ for the relation $v_\mathrm{max}$ vs.\ $T$. In addition, considering the GOES 1--8~{\AA} soft X-ray peak flux as an additional indicator for the thermal energy content of the flares, we find that $v_\mathrm{max}$ correlates better with the GOES peak flux ($c= 0.62$) than with the RHESSI $T$ and EM. For the relation between CME acceleration $a_\mathrm{max}$ and the GOES peak flux, the correlation coefficient is smaller, $c= 0.41$.

\begin{figure}[tbp]
\centering
\resizebox{7cm}{!}{\includegraphics{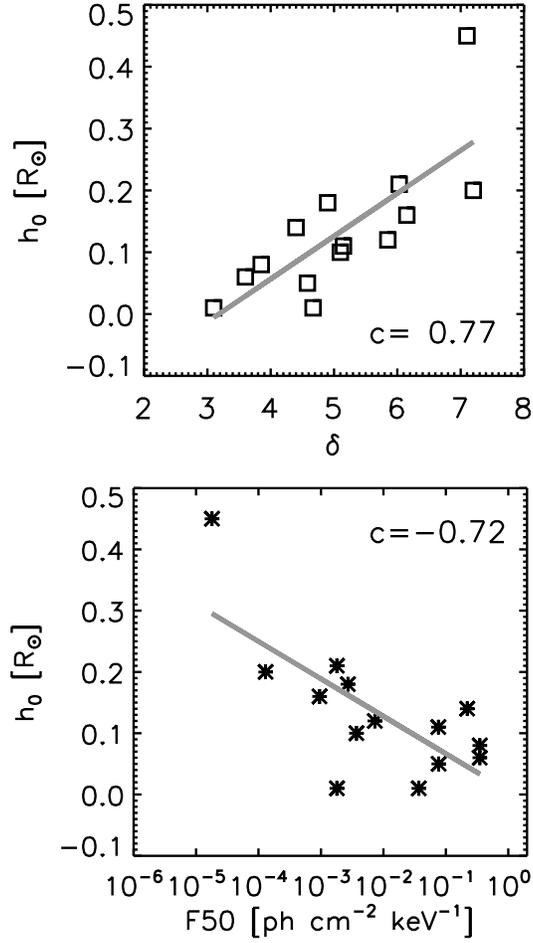}}
\caption{Scaling of the CME initiation height $h_0$ with the electron spectral index $\delta$ (top panel) and the flare X-ray flux at 50~keV, $F50$ (bottom panel) derived during the peak of the associated flares.}\label{fig4}
\end{figure}

We also obtained high correlations between the height $h_0$ above the solar surface at which a CME was observed for the first time, which can be interpreted as a measure of the initiation height of the pre-eruptive structure, and the non-thermal flare parameters (see Fig.~\ref{fig4}). The CME initiation height $h_0$ shows a high positive correlation with the spectral index $\delta$ of flare-accelerated electrons ($c= 0.77$) and a high inverse correlation with the non-thermal X-ray flux $F50$ ($c = -0.72$). This means that CMEs erupting at low coronal heights, i.e.\ in regions of stronger magnetic fields, are associated with flares in which a larger number of electrons is accelerated to high energies. 

The other CME height parameters we measured, $h_\mathrm{vmax}$ and $h_\mathrm{amax}$, i.e.\ the heights at which the CME velocity and CME acceleration reached their maximum, respectively, showed only weak or no correlations at all with the derived flare parameters. The highest correlation coefficient was obtained for the relation $h_\mathrm{vmax}$ and HXR duration $t_\mathrm{HXR}$ ($c= 0.47$), i.e.\ long duration events reach their peak velocity further out in the corona. We also compared the CME acceleration duration $t_\mathrm{acc}$ with the RHESSI spectral fit parameters, revealing no distinct relation except a weak correlation between $t_\mathrm{acc}$ and $\delta$ with $c= 0.41$. Consequently, CMEs with longer acceleration duration (and thus preferentially smaller peak acceleration, see \cite{bein2011}) show some tendency to be accompanied by flares with softer HXR spectra.


\subsection{Relative Timing of CME dynamics and Flare Energy Release}\label{sec:timing}

The first and second derivatives of the obtained CME height-time curves provided us with the times where the CME reached its maximum velocity and its maximum acceleration, as well as with the start and end time of the main CME acceleration phase. For each event, we derived the time difference of the start of the CME acceleration and the start of the non-thermal HXR emission of the associated flare as well as the time difference between the peak of the CME acceleration and the peak of the non-thermal HXR flare emissions, which marks the instant of the strongest particle acceleration. 

\begin{figure*}[tbp]
\centering
\resizebox{6.5cm}{!}{\includegraphics{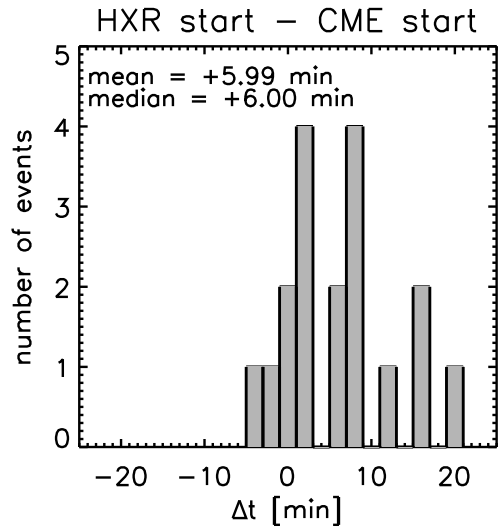}}
\caption{Distribution of the time lags between the start of the flare HXR emission and the start of the CME acceleration. Positive (negative) time lags indicate that the flare starts after (before) the CME acceleration.}\label{fig7}
\end{figure*}

In Fig.~\ref{fig7}, we show the distribution of the time lags obtained between the start of the flare HXR emission and the start of the CME acceleration. 
We find that in 83\% of the events the CME acceleration starts {\it before} the flare HXR emission. The distribution gives a mean of $+6.0\pm 9.0$~min and a median of $+6.0\pm 6.5$ min. 

\begin{figure*}[tbp]
\centering
\resizebox{6.5cm}{!}{\includegraphics{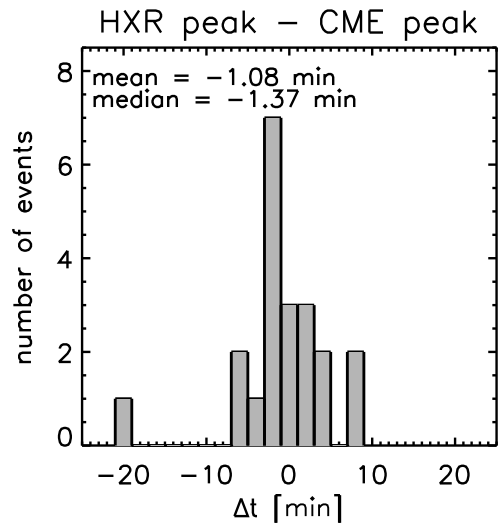}}
\caption{Distribution of the time lags between the peak of the flare HXR emission and the time of maximum CME acceleration.
Positive (negative) time lags indicate that the flare peaks after (before) the CME acceleration.
}\label{fig8}
\end{figure*}

Fig.~\ref{fig8} shows the distribution of the time difference between the peak of the flare HXR emission and the peak of the CME acceleration. We find that the maximum CME acceleration $a_\mathrm{max}$ occurs well synchronized with the flare HXR peaks. The arithmetic mean of the time lag distribution gives  $-1.1\pm 5.7$~min, the median $-1.4 \pm 2.2$~min.
In all but one case the time lags lie within in an interval of $[-10,+10]$~min. In $\sim$75\% of the CME-flare events under study, the flare HXR peak and the CME acceleration peak occur within 5~min of each other -- a time range, which corresponds to the typical uncertainties in the obtained CME acceleration peak times (cf.\ the shaded areas in Figs.~\ref{fig1}--\ref{fig3-1}). 
For comparison, the mean CME acceleration duration in the events under study is about 25~min. 

\begin{figure*}[tbp]
\centering
\resizebox{7cm}{!}{\includegraphics{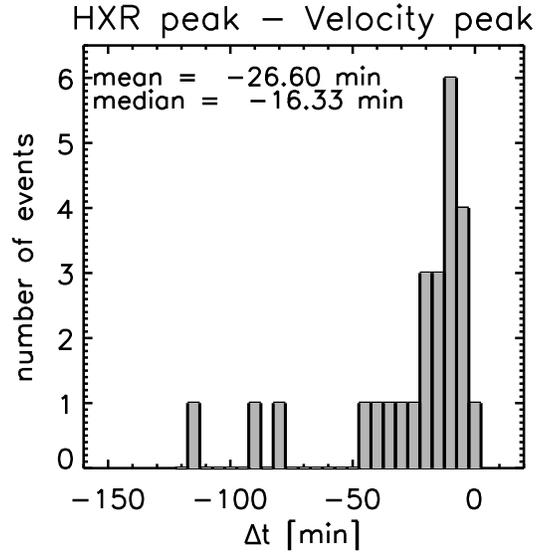}}
\caption{Distribution of the time lags between the peak of the flare HXR emission and the time of maximum CME velocity.
Positive (negative) time lags indicate that the flare peaks after (before) the CME reaches its maximum velocity.
}\label{fig9}
\end{figure*}

In Fig.~\ref{fig9}, we plot the distribution of the time lags between the peak of the flare HXR emission and the time when the CME reached its maximum velocity. 
We find that the CME velocities always reach their maximum {\it after} the HXR peak. The derived time difference $\Delta t$ lies in the range 2--117~min, with the 
median of the distribution at $-16.3\pm 8.5$~min.

\section{Discussion and Conclusions}\label{sec:discussion}

We investigated the physical relation between coronal mass ejections and their associated flares using several approaches. On the one hand, we determined the correlation and scaling of various parameters characterizing the CME acceleration with the flare's X-ray spectral parameters, which yield information on accelerated electrons as well as on the state of the thermal flare plasma. On the other hand, we studied the temporal relation between the CME acceleration and the flare energy release as evidenced in the non-thermal HXR radiation. 

Our results reveal a tight coupling between both phenomena. The CME peak velocity and peak acceleration yield distinct scalings with the flare parameters characterizing the accelerated electron spectra, in terms of the total number $e^-$ of accelerated electrons, the power in electrons $P20$, the HXR flux $F50$ at 50 keV, and the spectral index~$\delta$ of the electron spectra, with correlation coefficients in the range of 0.5 to 0.8 (all significant at least on the 95\% level). This means that CMEs with higher peak velocity and higher peak acceleration are accompanied by flares in which more electrons are accelerated, and in which a larger fraction of electrons is accelerated to higher energies (as it is revealed by the harder X-ray power-law spectra). The highest correlation coefficient in this study ($c = 0.85$) was obtained for the relation of the CME peak velocity $v_\mathrm{max}$,
which (together with its mass) determines the kinetic energy of the CME, and the product of the power in electrons above 20~keV and the duration of the  HXR emission, $P20\cdot t_\mathrm{HXR}$, which is a measure of the total energy in flare-accelerated electrons. These findings strongly support the general idea that the acceleration of the CME and the particle acceleration in the associated flare draw their energy from a common source, probably magnetic reconnection occurring in the current sheet behind the erupting structure. 

In general, the CME peak velocity is somewhat better correlated with the non-thermal flare parameters than the CME peak acceleration. However, there is one exception: the hardness of the accelerated electron spectrum yields a better correlation with the CME peak acceleration ($cc \approx -0.6$) than with the CME peak velocity ($cc \approx -0.5$), indicating that the hardness of the accelerated electron spectrum injected into the flare loops is intimately coupled to the impulsive acceleration process of the rising CME structure.  

We also found a distinct correlation of the CME initiation height $h_0$ and the spectral index $\delta$ of the flare-accelerated electrons ($c\approx 0.8$), as well as a distinct anti-correlation between $h_0$ and the non-thermal photon flux $F50$ ($c \approx -0.7$). We note that statistical studies of the CME main acceleration found an anticorrelation between the CME peak acceleration and the size and/or height of the pre-eruptive structure, with correlation coefficients of about $c \approx -0.5$ \citep{vrsnak2007,bein2011}. This  anticorrelation has been interpreted in terms of the Lorentz force driving the CME eruption and the variation of the coronal magnetic field strength with height: CMEs originating at low coronal heights, i.e. regions of stronger magnetic fields, have larger Lorentz forces available and can thus reach larger acceleration values than CMEs originating from high in the corona  where the magnetic field is smaller due to the (exponentially) decaying gas pressure and the related expansion of the magnetic field lines.  

Thus, the correlation between the hardness~$\delta$ of the flare electron spectrum and the CME initiation height~$h_0$ might be a secondary effect caused by the anticorrelation between the CME peak acceleration $a_{\rm max}$ and its initiation height~$h_0$. However, the correlation between flare $\delta$ and CME $h_0$ ($c \approx 0.8$) is significantly higher than that between 
CME $a_{\rm max}$ and CME $h_0$ \cite[$|c| \approx 0.5$;][]{bein2011}. We also stress that the initiation height $h_0$ is the CME parameter that gives the highest correlation coefficient with the hardness $\delta$ of the accelerated flare electron spectrum. These findings suggest that the height~$h_0$ of the pre-eruptive structure is a decisive parameter for the efficiency of the associated flare to accelerate electrons to high energies. 

The correlation coefficients obtained between the thermal flare parameters  and the CME peak velocity and peak acceleration are significantly smaller ($c\lesssim 0.5$) than those obtained for the non-thermal parameters. The fact, that both EM and $T$ show lower correlations with the CME $v_\mathrm{max}$ and $a_\mathrm{max}$ can be interpreted as an effect of the thermal flare plasma being only a secondary product within the flare process. The hot coronal flare plasma is generally assumed to be created by chromospheric heating and evaporation induced by the flare-accelerated electrons \citep[e.g.][]{neupert1968,brown1973,veronig2005}, which are a primary product of the flare energy release process. 

Several previous studies revealed a distinct relation between the CME mean velocities and the associated flares' soft X-ray peak flux measured by GOES, which characterizes the thermal energy content of flares \citep{moon2002,burkepile2004}. We note that these studies incorporated flare-CME events over a larger spread in flare importance by incorporating also X-class events. A sample of $55$ CME-flare pairs in the study of \citet{moon2002} suggest a correlation coefficient of $c\approx 0.5$ for the relation between the time integrated GOES SXR flux and the CME velocity.
\citet{burkepile2004} estimated the kinetic energy of CMEs originating from close to the solar limb and found a
higher correlation with the flare soft X-ray peak flux ($c = 0.74$). 
For our event sample, the correlation coefficient for $v_\mathrm{max}$ vs.\ GOES peak flux is somewhere in the middle with $c\approx 0.6$. 
We can summarize that the results obtained in the present paper for the thermal flare plasma are qualitatively in line with previous studies, and that our findings suggest that the CME peak acceleration and velocity are stronger coupled to the particle acceleration in the associated flares than to the maximum thermal energy content of the flare plasma. 

The comparison of flare HXR flux evolution and the acceleration profile of the CME main acceleration shows that in $\sim 80\%$ of the events under study, the non-thermal flare emission starts {\it after} the CME acceleration, on average delayed by $\approx 6$~min.
This finding agrees with investigations of the flare SXR emission in relation to CME acceleration by \cite{maricic2007} and \cite{bein2012} who also found that for the majority of the events, the CME acceleration starts before the flare SXR emission. Such delay of the flare start with respect to the start of the main CME acceleration is well in line with the standard flare model, where the rising flux rope stretches the field lines underneath. At a certain instant, in the current sheet behind the erupting structure magnetic reconnection will set in \cite[e.g. due to tearing instability, when the height-to-width ratio exceeds a certain threshold;][]{furth1963}, causing the main flare energy release and acceleration of high-energy particles. 

\begin{figure*}[tbp]
\centering
\resizebox{7cm}{!}{\includegraphics{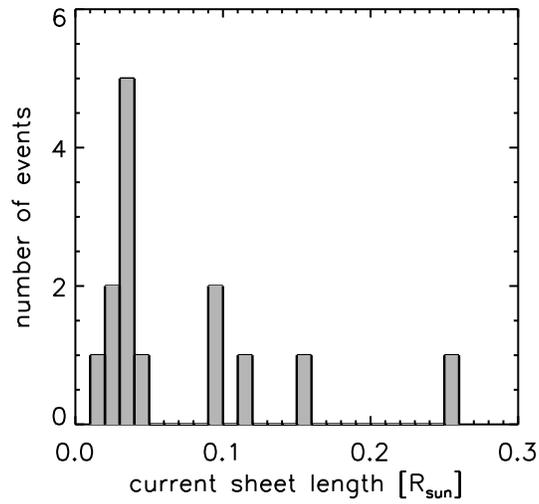}}
\caption{Distribution of current sheet length derived from the CME height at the onset of the non-thermal flare emission.
}\label{fig10}
\end{figure*}

Under these standard flare-CME model assumptions, we can estimate the length of the current sheet at the onset of magnetic reconnection.
For 14 flare-CME pairs in our sample, it was possible to derive the current sheet length from the CME height at the onset time of the  non-thermal HXR emission (i.e.\ particle acceleration) minus the initial height of the pre-eruptive structure; the distribution is plotted in Figure~\ref{fig10}. The median of the distribution indicates a current sheet length at the onset of magnetic reconnection of $0.03 \pm 0.01 R_{\odot}$, i.e. $21 \pm 7$~Mm in the events under study.  

The flare HXR peaks occur well synchronized with the peak of the CME acceleration profile. In 75\% of the cases they occur within $\pm5$~min, i.e. within the typical uncertainties in the determination of the CME accleration peak time. This means that at the time of the highest CME accleration also the rate of particle acceleration is highest. This finding agrees with the case studies by \citet{temmer2008,temmer2010} who also found a close synchronization of the flare HXR peak and the CME acceleration peak in  well observed limb events as well as fast halo CMEs. Other studies used the derivative of the flare SXR light curves to approximate the time evolution of the flare energy release, based on the Neupert effect \cite[e.g.][]{dennis1993,veronig2002}. For example, \citet{zhang2004} reported a close synchronization of the peak of the SXR flux derivative and the time of maximum acceleration in two long-duration CME-flare events. Statistically, 50--75\% of the events reveal a high degree of synchronization of the growth rate of SXR emission and CME acceleration, whereas about 25\% show strong deviations between the timing of the CME peak acceleration and the flare impulsive phase \citep{maricic2007,bein2012}.

To date, there exist no simulations of coupled CME-flare eruptions which directly incorporate particle acceleration mechanisms to
theoretically investigate the coupling of the CME dynamics and properties of accelerated flare particles. 
\citet{reeves2006} and \citet{reeves2010} performed MHD simulations of a flux rope eruption that leads to the formation of a large-scale current sheet and a multi-threaded flare beneath the CME, for which they calculated the thermal energy release and the expected rate of the flare SXR emission. They found that in cases where the background magnetic field and/or the magnetic reconnection rate is high, the CME acceleration and the associated thermal flare energy release are synchronized. Slow reconnection rates cause the CME acceleration to peak earlier, whereas for fast reconnection rates the acceleration peak shifted to later times in the eruption. The set of events we studied in this paper includes predominantly impulsive CMEs, characterized by high acceleration rates over a short acceleration duration. 
This selection, compared to the simulation results by \citet{reeves2010}, may explain why in our sample basically all events show a high synchronization of the peaks of the CME acceleration and the non-thermal flare emission. 

\acknowledgments
This activity has been supported by the European Community Framework Programme~7, High Energy Solar Physics Data in Europe (HESPE), grant agreement no.: 263086, the Austrian Space Applications Programme (ASAP-6 project \#819664 SOLDYN), and the Austrian Science Fund
(FWF): V195-N16.

\bibliographystyle{apj}

\end{document}